# Sequential Document Representations and Simplicial Curves

**Guy Lebanon**
Department of Statistics and
School of Electrical and Computer Engineering
Purdue University - West Lafayette, IN, USA

## Abstract

The popular bag of words assumption represents a document as a histogram of word occurrences. While computationally efficient, such a representation is unable to maintain any sequential information. We present a continuous and differentiable sequential document representation that goes beyond the bag of words assumption, and yet is efficient and effective. This representation employs smooth curves in the multinomial simplex to account for sequential information. In contrast to $n$-grams the new representation is able to robustly model long rage sequential trends in the document. We discuss the representation and its geometric properties and demonstrate its applicability for the task of text classification.

## 1 Introduction

Modeling of text documents is an essential part in a wide variety of applications such as the classification, segmentation, visualization and retrieval of text. A crucial part of the modeling process is choosing an appropriate representation for the documents. The original representation of documents as a sequence of words is extremely high-dimensional, discrete, and brittle making it an inconvenient representation for most models to work on. Instead, the words sequences are typically preprocessed by mapping them to a lower dimensional lossy representation such as $n$-gram, on which the modeling is carried out.

The $n$-gram representation keeps track of the number of $n$ consecutive words of different types, called $n$-grams. In the case $n = 1$ this amounts to ignoring the order of the words in the document, keeping only the histogram of word occurrences. This specific case, also called bag of words (bow) representation, is the most frequent one used due to its relative robustness in sparse situations. Assuming that the word counts are normalized by the document length, the bow representation of a word sequence $y = \langle y_1, \ldots, y_N \rangle$ over a vocabulary of integers $V = \{1, \ldots, |V|\}$ is $x \in \mathbb{R}^{|V|}$, where $x_j = \frac{1}{N} \sum_{i=1}^{N} \delta_{y_i,j}$. Alternatively, to avoid zeros the bow representation is often defined using smoothed word counts

$$x_j = \frac{1}{Z} \sum_{i=1}^{N} (\delta_{y_i,j} + c), \quad c \geq 0 \qquad (1)$$

where $Z$ ensures normalization. The smoothed version (1) may also be derived as the maximum posterior estimate for multinomial model and Dirichlet prior and is equal to the non-smooth version for $c = 0$.

Clearly, the $n$-gram representation sacrifices sequential information in favor of reduced dimensionality. In the case of $n = 1$ or bow representation, all sequential information is lost. In the case of $1 < n \ll N$, frequently occurring word patterns are kept, but long range sequential information concerning the word sequence $y$ is lost. Moreover, there is no way to tie the occurrences of word patterns to the specific locations in the document that they occurred. In terms of the applications mentioned above, the improvement that results from increasing $n$ is largely disappointing due to the high dimensionality and discreteness of the $n$-gram representation.

In this paper, we present a new representation called locally weighted bag of words (lowbow). This representation generalizes bow in a robust way that maintains long and short range sequential information. In contrast to $n$-gram that keeps track of frequently occurring patterns independent of their positions, lowbow keeps track of changes in the word histogram as it sweeps through the document from beginning to end.

One major advantage of lowbow is that it is a continuous and differentiable representation equivalent to smooth curves in the multinomial simplex. By naturally incorporating a smoothing kernel we are able to use lowbow effectively, even in cases that $n$-gram performs poorly due to its high dimensionality. Moreover, the smooth representation facilitates the use of tools from differentiable calculus and geometry to characterize the sequential contents of documents.

In the next section we describe the multinomial simplex and its geometry and then proceed with a formal definition of the lowbow representation.

## 2 The Multinomial Simplex and its Geometry

In this section, we present a brief description of the multinomial simplex and its geometry. Since the simplex is the space of bow representations, its geometry is crucial to the lowbow representation. The brief description below uses some concepts from Riemannian geometry. For additional information concerning the geometry of the simplex refer to (Amari & Nagaoka, 2000; Lafferty & Lebanon, 2005). A standard textbook on Riemannian geometry is (Spivak, 1975).

The multinomial simplex $\mathbb{P}_m$ for $m > 0$ is the $m$-dimensional subset of $\mathbb{R}^{m+1}$ of all positive probability vectors or histograms over $m+1$ objects

$$\mathbb{P}_m = \left\{ \theta \in \mathbb{R}^{m+1} : \forall i \ \theta_i > 0, \sum_{j=1}^{m+1} \theta_j = 1 \right\}.$$

Its connection to the multinomial distribution is that every $\theta \in \mathbb{P}_m$ corresponds to a multinomial distribution over $m+1$ items. The definition above does not include zero probabilities which lie on the boundary of the simplex closure $\overline{\mathbb{P}_m}$. As a result, $\mathbb{P}_m$ is a differentiable manifold and we can use the standard language of differentiable manifolds, rather than the more complicated notion of a geometry with corners. A discussion concerning the positivity restriction and its relative unimportance in practice may be found in (Lafferty & Lebanon, 2005).

The topological structure of $\mathbb{P}_m$, which determines the notions of convergence and continuity, is naturally inherited from the standard topological structure of the embedding space $\mathbb{R}^{m+1}$. The geometrical structure of $\mathbb{P}_m$ that determines the notions of distance, angle and curvature is determined by a local inner product $g_\theta(\cdot, \cdot), \theta \in \mathbb{P}_m$, called the Riemannian metric. The most obvious choice, perhaps, is the standard Euclidean inner product $g_\theta(u, v) = \sum u_i v_i$. However, such a choice is problematic from several aspects. Fortunately, a particular choice of a Riemannian metric is motivated by Čencov (1982) who proved that the

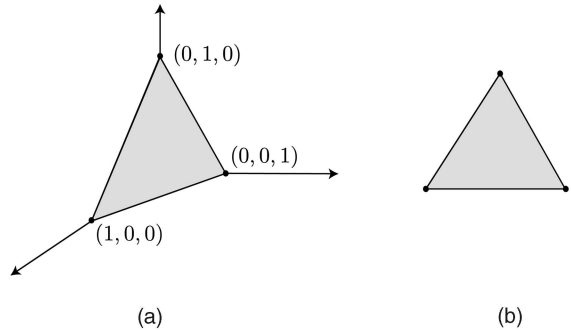

Figure 1: The 2-simplex $\mathbb{P}_2$ may be visualized as a surface in $\mathbb{R}^3$ (left) or as a triangle in $\mathbb{R}^2$ (right).

Fisher information metric

$$g_\theta(u, v) = \sum_{ij} u_i v_j \mathsf{E}_{p_\theta} \left( \frac{\partial \log p_\theta(x)}{\partial \theta_i} \frac{\partial \log p_\theta(x)}{\partial \theta_j} \right) \quad (2)$$

$$= \sum_{i=1}^{m+1} \frac{u_i v_i}{\theta_i} \quad (3)$$

(above, $p_\theta(x)$ is the multinomial probability associated with the parameter $\theta$) is the only invariant metric under sufficient statistics transformations. In addition, various recent results motivate the Fisher geometry from a practical perspective (Lafferty & Lebanon, 2005).

The inner product (3) defines the geometric properties of distance, angle and curvature on $\mathbb{P}_m$ in a way that is quite different from the Euclidean inner product. The distance function $d : \mathbb{P}_m \times \mathbb{P}_m \to [0, \pi/2]$ corresponding to (3) is

$$d(\theta, \eta) = \mathrm{acos} \left( \sum_{i=1}^{m+1} \sqrt{\theta_i \eta_i} \right). \quad (4)$$

The distance function (4) and the Euclidean distance function $d(\theta, \eta) = \sqrt{\sum (\theta_i - \eta_i)^2}$ resulting from the inner product $g_\theta(u, v) = \sum u_i v_i$ on $\mathbb{P}_2$ are illustrated in Figures 1-2.

By determining geometric properties such as distance, the choice of metric for $\mathbb{P}_m$ is of direct importance to the bow representation of documents and its modeling. For example, while the Euclidean metric is homogenous across the simplex, the Fisher metric (3) emphasizes the area close to the boundary. The next section describes the lowbow representation which amounts to a parameterized curve in $\mathbb{P}_m$. In addressing the question of modeling such curves, the simplex geometry plays a central role. It dictates notions such as the distance between two curves, the instantaneous direction of a curve, and the curvature or complexity of a curve. Understanding the relationship between these geometric notions and $g_\theta$ is the first step in modeling documents using the lowbow representation.

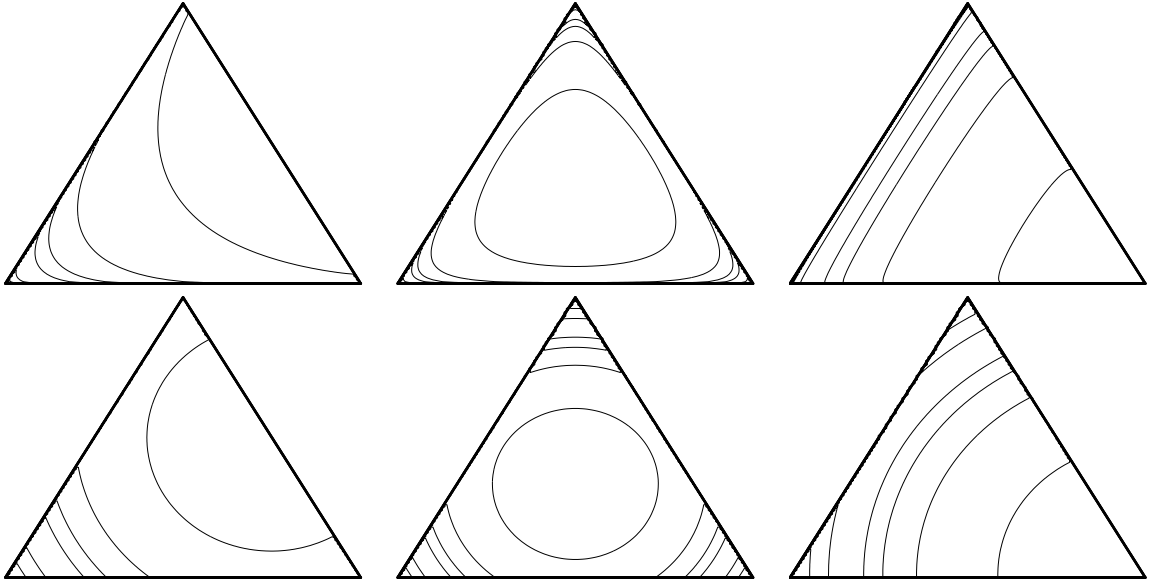

Figure 2: Equal distance contours on $\mathbb{P}_2$ from the upper right edge (left column), the center (center column), and lower right corner (right column). The distances are computed using the Fisher information metric (top row) and the Euclidean metric (bottom row).

## 3 Sequential Document Representations

We typically think of documents over a vocabulary $V = \{1, \ldots, |V|\}$ as a finite sequence $\langle y_1, \ldots, y_N \rangle$ of words represented as integers in $V$. However, we use the following broader definition for documents that will be more convenient later on.

**Definition 1.** *A document $x$ of length $N$ is a function $x : \{1, \ldots, N\} \times V \to [0, 1]$ such that*

$$\sum_{j=1}^{|V|} x(i, j) = 1 \quad \forall i \in \{1, \ldots, N\}.$$

*The set of documents (of all lengths) is denoted by $\mathfrak{X}$.*

For a document $x \in \mathfrak{X}$ the value $x(i, j)$ is the weight of the word $j \in V$ at location $i$. The standard way to represent a word sequence as a document in $\mathfrak{X}$ is to have each location host the appropriate single word with constant weight, which corresponds to the $\delta_c$ representation defined below with $c = 0$.

**Definition 2.** *The standard representation $\delta_c(y) \in \mathfrak{X}$, where $c \geq 0$, of a word sequence $y = \langle y_1, \ldots, y_N \rangle$ is*

$$\delta_c(y)(i, j) = \begin{cases} \frac{c}{1+c|V|} & y_i \neq j \\ \frac{1+c}{1+c|V|} & y_i = j \end{cases}.$$

The above is a legitimate representation since $\sum_j \delta_c(y)(i, j) = \frac{1+c|V|}{1+c|V|} = 1$. The parameter $c$ in the above definition is useful for avoiding zero counts in the obtained local histograms $\delta_c(y)(i, \cdot)$.

The standard representation $\delta_c$ assumes that each word $y_j$ in the sequence $y$ occupies a single location $1, \ldots, N$. In general, however, Definition 1 lets several words occupy the same location by smoothing the influence of words $y_j$ across different document positions. Doing so is central in converting the discrete-time standard representation to a continuous representation that is much more convenient for analysis and modeling.

We are interested in comparing different documents on the basis of their sequential contents. Definition 1 is problematic since it depends on the length of the document. We would like to treat two documents with similar sequential contents but different lengths in a similar fashion. Towards this goal we introduce length-normalized documents that abstract away from the document length by normalizing the sequential information to lie in the interval $[0, 1]$.

**Definition 3.** *A length-normalized document $x$ is a function $x : [0, 1] \times V \to [0, 1]$ such that*

$$\sum_{j=1}^{|V|} x(t, j) = 1, \quad \forall t \in [0, 1].$$

*The set of length-normalized documents is denoted $\mathfrak{X}^*$.*

The procedure of converting a document $x \in \mathfrak{X}$ to a length-normalized document $x' \in \mathfrak{X}^*$ is expressed by the length-normalization function defined below.

**Definition 4.** *The length-normalization function for a document $x \in \mathfrak{X}$ of length $N$ is*

$$\varphi : \mathfrak{X} \to \mathfrak{X}^* \quad \varphi(x)(t,j) = x(\lceil tN \rceil, j)$$

*where $\lceil r \rceil$ is the smallest integer greater than $r$.*

With each length-normalized document we can associate the following global bow or histogram.

**Definition 5.** *The global bow representation of a length-normalized document $x \in \mathfrak{X}^*$ is*

$$\rho : \mathfrak{X}^* \to \mathbb{P}_{|V|-1} \quad [\rho(x)]_j = \int_0^1 x(t,j)\,dt.$$

Note that the function $\rho$ is well defined since

$$\sum_{j=1}^{|V|}[\rho(x)]_j = \sum_{j=1}^{|V|}\int_0^1 x(t,j)\,dt = \int_0^1 \sum_{j=1}^{|V|} x(t,j)\,dt$$

$$= \int_0^1 1\,dt = 1 \implies \rho(x) \in \mathbb{P}_{|V|-1}.$$

As proved in Theorem 2, the global bow representation defined above for a document in its standard representation $\delta_c(y)$ is equivalent to the popular definition of bow expressed in equation (1).

We describe below a smoothed representation in $\mathfrak{X}^*$ that is analogous to a curve in the multinomial simplex. The smoothed representation is obtained by convolving the length-normalized standard representation $\varphi(\delta_c(y))$ with a positive smoothing kernel. A positive smoothing kernel is a function $K_{\mu,\sigma} : [0,1] \to (0,\infty)$, parameterized by location and scale parameters $\mu \in [0,1], \sigma \in (0,\infty)$. The parameter $\mu$ represents the (normalized) document location and $\sigma$ represents the amount of smoothing. We further assume that a smoothing kernel is continuous and differentiable in $\mu, r$ and normalized, i.e., $\int_0^1 K_{\mu,\sigma}(t)\,dt = 1$.

One example for a smoothing kernel is the normal distribution pdf restricted to $[0,1]$ and normalized

$$K_{\mu,\sigma}(x) = \begin{cases} \frac{1}{Z} N(x\,;\mu,\sigma) & x \in [0,1] \\ 0 & x \notin [0,1] \end{cases}. \quad (5)$$

Another example is the beta distribution pdf

$$K_{\mu,\sigma}(x) = \text{Beta}\left(x\,;\ c\frac{\mu}{\sigma}\,,\ c\frac{1-\mu}{\sigma}\right) \quad (6)$$

where $c$ is selected so that the two parameters of the beta distribution will be greater than 1. The above Beta pdf has expectation $\mu$ and variance that is increasing in the scale parameter $\sigma$. The following definitions complete the description of the lowbow representation.

**Definition 6.** *The $\mu$-modulation function $\psi_\mu$ maps a document $x \in \mathfrak{X}^*$ to*

$$\psi_\mu(x)(t,j) = x(t,j)K_{\mu,\sigma}(t) \quad t \in [0,1], j \in V.$$

**Definition 7.** *The locally weighted bag of words (lowbow) representation at $\mu$ of the word sequence $y$ is*

$$\gamma_\mu(y) = \rho \circ \psi_\mu \circ \varphi \circ \delta_c(y) \in \mathbb{P}_{|V|-1}, \quad \mu \in [0,1].$$

First note that lowbow indeed maps a word sequence $y$ and $\mu$ into the simplex $\gamma_\mu(y) \in \mathbb{P}_{|V|-1}$ since $\sum_j [\gamma_\mu(y)]_j = \sum_j \int_0^1 x(t,j) K_{\mu,\sigma}(t)\,dt = \int_0^1 K_{\mu,\sigma}(t) \sum_j x(t,j)\,dt = \int_0^1 K_{\mu,\sigma}(t) \cdot 1\,dt = 1$.

**Theorem 1.** *The lowbow representation is a continuous and differentiable parameterized curve in the simplex, in both the Euclidean and the Fisher geometry.*

*Proof.* We prove below only the continuity of the lowbow representation. The proof of differentiability proceeds along similar lines. Fixing $y$, the mapping $\mu \mapsto \gamma_\mu(y)$ maps $[0,1]$ into the simplex $\mathbb{P}_{|V|-1}$. Since $K_{\mu,\sigma}(t)$ is continuous on a compact region $(\mu,t) \in [0,1]^2$, it is also uniformly continuous and we have

$$\lim_{\epsilon \to 0} |[\gamma_\mu(y)]_j - [\gamma_{\mu+\epsilon}(y)]_j|$$

$$= \lim_{\epsilon \to 0} \Big| \int_0^1 x(t,j)K_{\mu,\sigma}(t) - x(t,j)K_{\mu+\epsilon,\sigma}(t)dt \Big|$$

$$\leq \lim_{\epsilon \to 0} \int_0^1 x(t,j)|K_{\mu,\sigma}(t) - K_{\mu+\epsilon,\sigma}(t)|\,dt$$

$$\leq \lim_{\epsilon \to 0} \sup_{t \in [0,1]} |K_{\mu,\sigma}(t) - K_{\mu+\epsilon,\sigma}(t)| \int_0^1 x(t,j)\,dt$$

$$\leq \lim_{\epsilon \to 0} \sup_{t \in [0,1]} |K_{\mu,\sigma}(t) - K_{\mu+\epsilon,\sigma}(t)| = 0.$$

As a result,

$$\lim_{\epsilon \to 0} \|\gamma_\mu(y) - \gamma_{\mu+\epsilon}(y)\|$$

$$= \sqrt{\sum |[\gamma_\mu(t)]_j - [\gamma_{\mu+\epsilon}(t)]_j|^2} \to 0$$

proving the continuity of $\gamma_\mu(y)$ in the Euclidean geometry. Since the Euclidean geometry and the Fisher geometry share the same topology, the lowbow curves are also continuous in the Fisher geometry. $\square$

It is important to note that the parameterized curve that corresponds to the lowbow representation consists of two parts: the geometric figure $\{\gamma_\mu(y) : \mu \in [0,1]\} \subset \mathbb{P}_{|V|-1}$ and the parameterization function $\mu \mapsto \gamma_\mu(y)$ that ties the local histogram to a location $\mu$ in the document. A common mistake in dealing with parameterized curves is to focus on the geometric figure while ignoring the parameterization. This is a mistake since different lowbow representations may share similar geometric figures but possess different parameterization speeds indicating different local properties.

The geometric properties of the curve depend on the word sequence, the kernel shape and the kernel scale parameter. The kernel scale parameter is especially important as it determines the amount of temporal smoothing employed. If $\sigma \to \infty$ the lowbow curve degenerates into a single point that is the global bow representation.

**Theorem 2.** *Let $K_{\mu,\sigma}$ be a smoothing kernel such that when $\sigma \to \infty$, $K_{\mu,\sigma}(x)$ is constant in $\mu, x$. Then for $\sigma \to \infty$, the lowbow curve $\gamma(y)$ degenerates into a single point equivalent to the bow representation (1).*

*Proof.* Since the kernel is both constant and normalized, $K_{\mu,\sigma}(t) = 1, \forall \mu, t \in [0,1]^2$. For all $\mu \in [0,1]$,

$$[\gamma_\mu(y)]_j = \int_0^1 \varphi(\delta_c(y))(t,j) K_{\mu,\sigma}(t)\, dt$$
$$= \int_0^1 \varphi(\delta_c(y))(t,j)\, dt$$
$$= \sum_{i=1}^N \frac{1}{N} \left( \delta_{y_i,j} \frac{1+c}{1+c|V|} + (1-\delta_{y_i,j}) \frac{c}{1+c|V|} \right)$$
$$\propto \sum_{i=1}^N \delta_{y_i,j}(1+c) + (1-\delta_{y_i,j})c \propto \sum_{i=1}^N (\delta_{y_i,j} + c).$$

□

In the other extreme, small $\sigma$ will result in a curve figure that quickly moves between the different corners of the simplex as the words $\langle y_1, \ldots, y_N \rangle$ are encountered at times $t = 1/N, \ldots, N/N$. Thus the lowbow representation as $\sigma \to 0$ approaches a representation equivalent to the original word sequence. It is unlikely that either extreme cases $\sigma \to 0$ or $\sigma \to \infty$ will be optimal from a modeling perspective. By varying $\sigma$ between 0 and $\infty$ we are able to interpolate between these two extreme cases and obtain a convenient sequential, yet robust, representation.

Figure 3 illustrates the geometric figure resulting from the lowbow representation and its dependency on the kernel scale parameter and the smoothing coefficient. Notice how the figure shrinks as $\sigma$ increases until it reaches the single point that is bow. Increasing $c$, on the other hand, pushes the geometric figure towards the center of the simplex.

It is useful to have a quantitative characterization of the complexity of the lowbow representation as a function of the chosen kernel and $\sigma$. Towards this end, the kernel's complexity, defined below, serves as a bound for variations in the lowbow curve.

**Definition 8.** *Let $K_{\mu,\sigma}(t)$ be a kernel that is Lipschitz continuous[1] in $\mu$ with a Lipschitz constant $C_K(t)$. The kernel's complexity is defined as*

$$\mathcal{O}(K) = \sqrt{|V|} \int_0^1 C_K(t)\, dt.$$

The theorem below proves that the lowbow curve is Lipschitz continuous with a Lipschitz constant $\mathcal{O}(K)$, thus connecting the curve complexity with the shape and the scale of the kernel.

**Theorem 3.** *The lowbow curve $\gamma(y)$ satisfies*
$$\|\gamma_\mu(y) - \gamma_\tau(y)\| \leq |\mu - \tau|\, \mathcal{O}(K).$$

*Proof.*
$$|[\gamma_\mu(y)]_j - [\gamma_{\mu+\epsilon}(y)]_j|$$
$$\leq \int_0^1 x(t,j)|K_{\mu,\sigma}(t) - K_{\mu+\epsilon,\sigma}(t)|\, dt$$
$$\leq \int_0^1 |K_{\mu,\sigma}(t) - K_{\mu+\epsilon,\sigma}(t)|\, dt \leq \epsilon \int_0^1 C_K(t)\, dt$$

and so $\|\gamma_\mu(y) - \gamma_{\mu+\epsilon}(y)\| \leq \epsilon \mathcal{O}(K)$. □

While Theorem 3 is expressed in terms of the Euclidean distance on the simplex, it is straightforward to extend it to the case of the Fisher geodesic distance $d(\gamma_\mu(y), \gamma_\tau(y))$ as it is a smooth function of the Euclidean distance.

## 4 Modeling of Simplicial Curves

A lowbow representation is equivalent to a parameterized curve in the simplex. As such, it is a point in an infinite product of simplices $\mathbb{P}_m^{[0,1]}$, that is naturally equipped with the product topology and geometry of the individual simplices. In practice, maintaining a continuous representation is difficult, and the straightforward solution is to sample the path at representative points $t_1, \ldots, t_l \in [0,1]$ resulting in a discrete lowbow representation equivalent to a point in the finite dimensional product space $\mathbb{P}_m^l$. We describe below geometrical features of lowbow and their use in modeling. The discussion focuses on continuous lowbow curves case, but the discrete case resulting from finite sampling may be obtained by replacing integrals with sums.

The distance between lowbow representations of two word sequences $\gamma(y_1), \gamma(y_2) \in \mathbb{P}_m^{[0,1]}$, is the average distance between the corresponding time coordinates

$$d(\gamma(y_1), \gamma(y_2)) = \int_0^1 d(\gamma_t(y_1), \gamma_t(y_2))\, dt \qquad (7)$$

---
[1] A Lipschitz continuous function $f$ satisfies $|f(x) - f(y)| \leq C|x - y|$ for some constant $C$ called the Lipschitz constant.

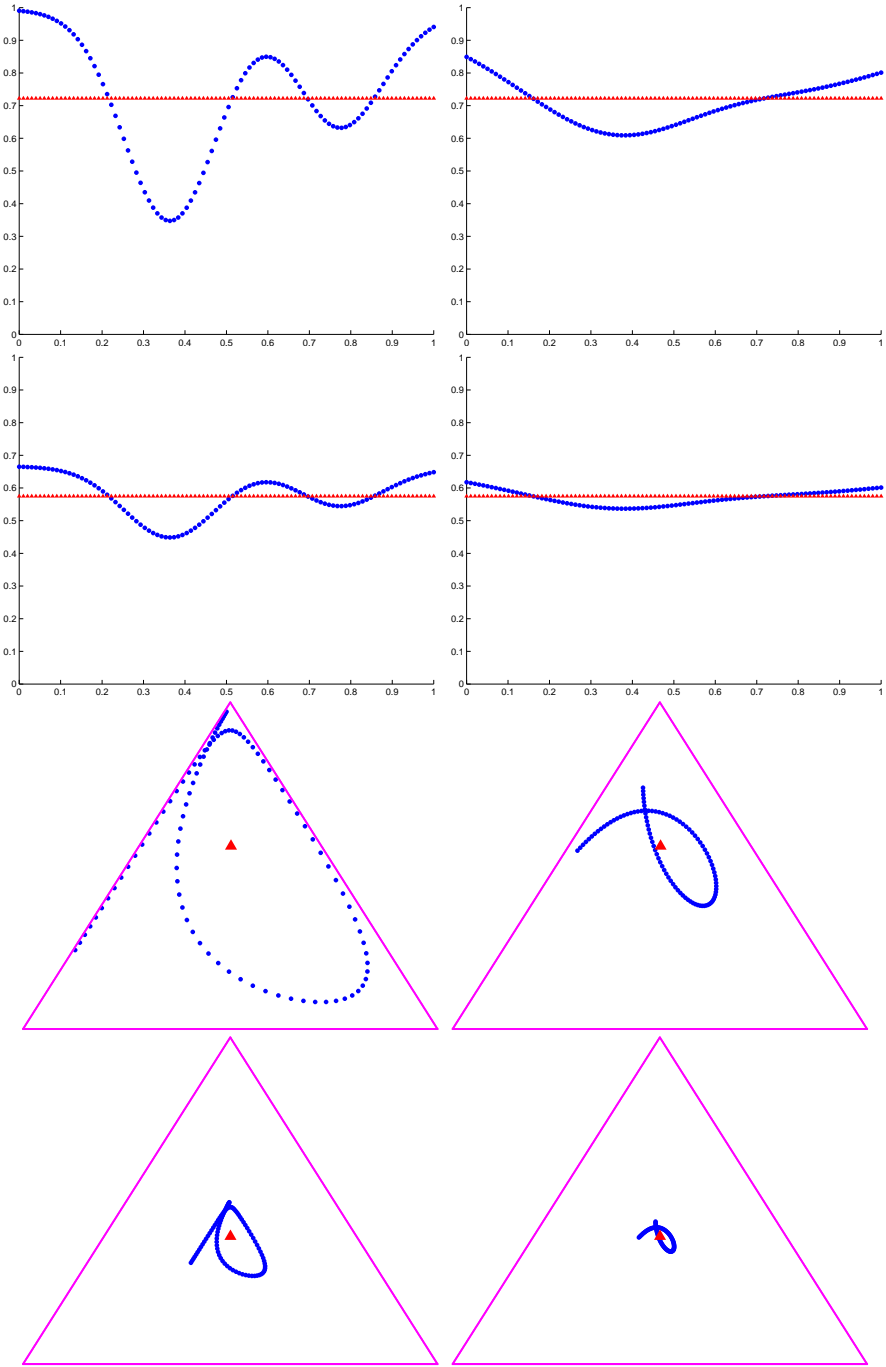

Figure 3: The geometric figure resulting from the lowbow representation of the word sequences ⟨1 1 1 2 2 1 1 1 2 1 1⟩ (which is a curve in $\mathbb{P}_1$ – top two rows) and ⟨1 3 3 3 2 2 1 3 3⟩ (which is a curve in $\mathbb{P}_2$ – bottom two rows). The figures illustrate the differences as the kernel scale parameter increases from 0.1 to 0.2 (left vs. right column) and the smoothing coefficient $c$ varies from 0.005 to 1 (first and third rows vs. second and fourth rows). Increasing the kernel scale causes some local features to vanish, for example the local mode in the first word sequence and the tail in the bottom left corner of $\mathbb{P}_2$ in the second word sequence. In addition, increasing $\sigma$ shrinks the figure towards the single bow point (represented by the horizontal line in the top two rows and the triangle in the bottom two rows). Increasing the smoothing coefficient $c$ causes the figure to stay away from the boundary of the simplex and concentrate in the center. Since the curves are composed of 100 dots, the distances between the dots indicate the parameterization speed of the curves.

where $d(\gamma_t(y_1), \gamma_t(y_2))$ depends on the simplex geometry under consideration, e.g. Equation (4) or the Euclidean norm.

The distance between lowbow curves $d(\gamma(y_1), \gamma(y_2))$ (or its discrete analog) may be used in various modeling tasks. In $k$-nearest neighbors it simply replaces traditional bow-based distances such as tf similarity or the Euclidean norm. The distance (7) may also be used to construct the approximated heat kernel for use in SVM classification and regression as described by Lafferty and Lebanon (2005). The lowbow distance may also be used to construct generative models for text that generalizes the naive Bayes or multinomial model. Such language models serve an important role in applications such as machine translation, speech recognition and information retrieval.

Other geometrical features may also be used in modeling text documents. The instantaneous direction of the curve $\gamma(y)$ at $t_0$ is given by its tangent vector $\dot{\gamma}_{t_0}(y)$. Since the curve is differentiable we can obtain a tangent vector field $\dot{\gamma}(y)$ along the curve that describes sequential topic trends and their change. The second derivative $\ddot{\gamma}(y)$ vector field, together with the Riemannian metric, may be used to define the curvature at different points along the lowbow curve. Intuitively, curvature measures the amount of wigglyness or deviation from a straight line (or geodesic). Integrating the norm of the curvature tensor over $t \in [0, 1]$ provides a measure of the sequential topic complexity or variability along the document. The tangent vector field and curvature of lowbow may be used to study local properties of documents that relate to tasks such as segmentation or summarization.

The geometric properties mentioned above are useful in constructing generative and conditional models for text applications such as retrieval, classification, filtering, segmentation and visualization. Due to lack of space we will briefly present in the next section a simple lowbow $k$-nearest neighbor classifier. The application of lowbow to additional tasks, and other classification models will be studied in future work.

## 5 Text Classification Experiments

In this section, we examine lowbow and its properties in the context of text classification using a nearest neighbor classifier. We report experimental results for the WebKB faculty vs. course task and the Reuters top ten categories using the standard mod-apte split. In the WebKB task we repeatedly sampled subsets for training and testing with equal positive and negative examples. In the Reuters task we randomly sampled subsets of the mod-apte split for training (to examine the influence of the train set size) resulting in unbalanced train and test set containing more negative than positive examples. The continuous quantities in the lowbow calculation were approximated by a discrete sample of 5 equally spaced points in the interval $[0, 1]$ turning the integrals into efficiently computed Riemann sums. The kernel used was the bounded Gaussian kernel (5). Throughout the experiments we computed several alternatives for the kernel scale parameter and chose the best one. An important extension of the current experiments that we plan to conduct in the future is automatic selection of $\sigma$.

Figure 4 (top) displays results for the WebKB data. The left graph is a standard train-set size vs. test set error rate comparing the lowbow geodesic using scale $\sigma \to \infty$ (dashed) and the lowbow geodesic distance using an intermediate scale value. The right graph display the dependency of the test set error rate on the scale parameter indicating an optimal scale at around $\sigma = 0.2$ (for repeated samplings of 500 training examples). In both cases, the performances of standard bow techniques such as tf cosine similarity or Euclidean distance were significantly inferior (20-40% higher error rate) than the displayed curves and are not displayed.

Figure 4 (bottom) displays test set error rates for the Reuters task. The 10 rows in the table indicate the classification task of identifying each of the 10 most popular classes in the Reuters collection. The columns represent varying training set sizes (sampled from the mod-apte split). The lowbow geodesic distance for an intermediate scale is denoted by $\text{err}_1$ and for $\sigma \to \infty$ is denoted by $\text{err}_2$. tf-Cosine similarity and Euclidean distance for bow are denoted by $\text{err}_3$ and $\text{err}_4$.

The experiments indicate that lowbow geodesic clearly outperforms, for most values of $\sigma$, the standard tf-cosine similarity and Euclidean distance for bow. In addition they also indicate that in general, the best scale parameter for lowbow is an intermediate one - and not $\sigma \to \infty$ thus validating the hypothesis that we can leverage sequential information using the lowbow framework to improve on global bow models. Whether or not the optimal value of $\sigma$ is small or large depends on several factors. It depends on the relationship between the sequential information and the response variable as some tasks are more amenable to such an approach than others. The size of the training set matters as well since excessive lowbow smoothing will be required for extremely small datasets resulting in an optimal scale of $\sigma \to \infty$.

## 6 Related Work and Conclusion

The use of $n$-gram and bow in document modeling has a long history. A geometric point of view considering the bow representation as a point in the multinomial simplex is expressed in (Lafferty & Lebanon, 2005), and a recent overview of the geometrical properties of probability spaces is (Amari & Nagaoka, 2000). The use of simplicial curves in text modeling is a relatively new approach promoted by Gous (1998) and Hall and Hofmann (). **?** (**?**) describes some related ideas that lead to a non-smooth multi-scale view of images.

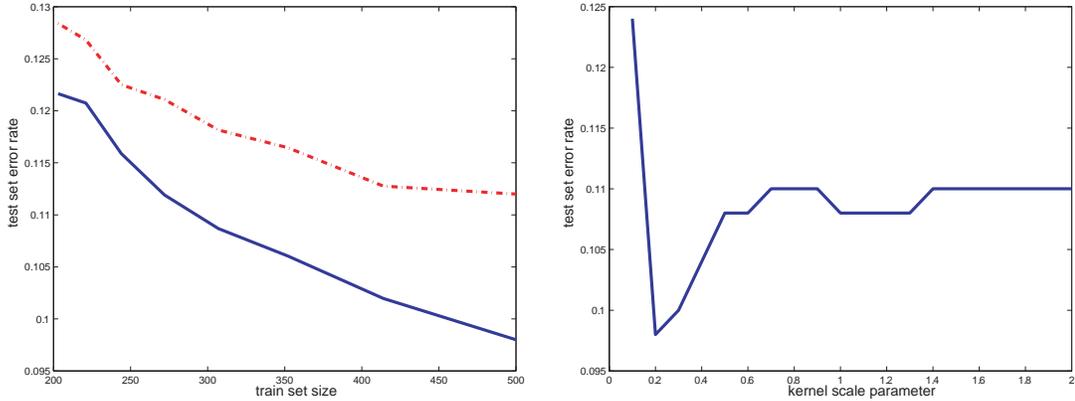

| | Train Size = 100 | | | | Train Size = 200 | | | | Train Size = 400 | | | |
|---|---|---|---|---|---|---|---|---|---|---|---|---|
| class | $\text{err}_1$ | $\text{err}_2$ | $\text{err}_3$ | $\text{err}_4$ | $\text{err}_1$ | $\text{err}_2$ | $\text{err}_3$ | $\text{err}_4$ | $\text{err}_1$ | $\text{err}_2$ | $\text{err}_3$ | $\text{err}_4$ |
| 1 | **9.9** | 10.6 | 11.2 | 11.0 | **8.1** | 9.7 | 8.2 | 10.7 | **6.7** | 7.3 | 11.2 | 9.0 |
| 2 | **11.6** | 12.8 | 17.6 | 22.4 | **9.4** | 9.7 | 17.6 | 19.9 | 7.9 | **7.8** | 17.2 | 17.9 |
| 3 | **6.8** | 7.6 | 6.9 | 12.9 | **5.8** | 7.2 | 7.8 | 16.9 | 5.4 | **5.3** | 10.2 | 12.6 |
| 4 | 5.6 | 6.5 | 6.5 | **5.5** | **4.8** | **4.8** | 7.1 | 7.0 | **4.5** | 4.7 | 8.5 | 7.5 |
| 5 | 6.6 | **6.2** | 9.0 | 11.4 | **5.7** | 6.8 | 6.7 | 10.3 | **5.0** | 5.6 | 5.8 | 7.4 |
| 6 | **5.7** | 5.8 | 5.8 | 10.8 | **5.2** | 5.3 | 5.3 | 10.0 | **4.8** | 5.4 | 5.6 | 11.3 |
| 7 | **4.2** | 5.1 | 7.0 | 12.9 | **4.2** | 4.3 | 7.9 | 9.0 | **3.9** | 4.3 | 5.8 | 7.5 |
| 8 | **3.0** | 3.2 | 4.7 | 7.6 | **3.0** | 3.3 | 3.4 | 3.4 | **2.6** | 2.9 | 3.2 | 3.9 |
| 9 | **2.8** | 4.0 | 4.9 | 7.9 | 3.1 | **3.0** | 6.4 | 2.8 | **2.9** | 3.2 | 4.7 | 5.1 |
| 10 | 2.7 | 2.9 | 3.6 | **2.6** | **2.6** | 3.0 | 5.8 | 3.1 | 2.3 | 2.6 | 3.7 | **2.2** |

Figure 4: Experimental test set error rates for WebKB course vs. faculty task (top) and Reuters top 10 classes using samples from mod-apte split (bottom). See Section 5 for details.

The lowbow representation is a promising new direction in text modeling. By varying $\sigma$ it interpolates between the standard word sequence representation $\langle y_1, \ldots, y_N \rangle$ and bow. It is equivalent to a smooth curve that is amenable to techniques from real analysis and geometry. In contrast to $n$-gram, it captures topical trends and incorporates long range information. On the other hand, the lowbow novelty is orthogonal to $n$-gram as it is possible to combine the two.

The lowbow framework is aesthetically pleasing, and achieves good results in practice. Using a smoothing parameter, it naturally interpolates between bow and complete sequential information. The correspondence with smooth curves in the simplex enables the use of a wide array of tools from analysis and geometry in an otherwise discrete representation. It seems likely that the lowbow representation and its geometric properties will lead to improvements in many areas of text modeling, including retrieval, classification, visualization, segmentation, summarization, and language modeling.